\documentclass{elsart}
\usepackage{latexsym,amsmath,tabularx,amssymb,bm}
\usepackage{epsfig}
\usepackage{citesort}
\usepackage{graphicx}
\usepackage{amssymb}
\usepackage{makeidx}
\usepackage{graphicx}
\usepackage{multicol}

\long\def\comment#1{ }

\newcommand{\dif}{{\rm d}}
\newcommand{\cal}{\mathcal}
\newcommand{\abar}{\bar{\alpha}_s}
\newcommand{\mean}[1]{\left\langle #1 \right\rangle_Y}
\newcommand{\means}[1]{\langle #1 \rangle_Y}

\newcommand{\beq}{\begin{eqnarray}}
\newcommand{\eeq}{\end{eqnarray}}
\newcommand{\be}{\begin{eqnarray}}
\newcommand{\ee}{\end{eqnarray}}

\def\simge{\mathrel{%
   \rlap{\raise 0.511ex \hbox{$>$}}{\lower 0.511ex \hbox{$\sim$}}}}
\def\simle{\mathrel{
   \rlap{\raise 0.511ex \hbox{$<$}}{\lower 0.511ex \hbox{$\sim$}}}}
\def\bigs{\mathrel{
   \rlap{\raise 0.531ex \hbox{$>$}}{\lower 0.531ex \hbox{$<$}}}}

\begin{document}
\begin{flushright}
~\vspace{-1.25cm}\\
{\small\sf SACLAY--T05/012}
\end{flushright}
\vspace{0.8cm}
\begin{frontmatter}

\parbox[]{16cm}{ \begin{center}
\title{Non--linear QCD evolution with improved triple--pomeron vertices}

\author{E.~Iancu\thanksref{th2}} and
\author{\ D.N.~Triantafyllopoulos}
\address{Service de Physique Th\'eorique, CEA/DSM/SPhT,  Unit\'e de recherche
associ\'ee au CNRS (URA D2306), CE Saclay,
        F-91191 Gif-sur-Yvette, France}

\thanks[th2]{Membre du Centre National de la Recherche Scientifique
(CNRS), France.}

\date{\today}
\vspace{0.8cm}
\begin{abstract}
In a previous publication, we have constructed a set of non--linear
evolution equations for dipole scattering amplitudes in QCD at high
energy, which extends the Balitsky--JIMWLK hierarchy
by including the effects of fluctuations in the gluon number
in the target wavefunction. In doing so, we have
relied on the color dipole picture, valid in the limit where the
number of colors is large, and we have made some further
approximations on the relation between scattering amplitudes and
dipole densities, which amount to neglecting the non--locality of
the two--gluon exchanges. In this Letter, we relax the latter
approximations, and thus restore the correct structure of the
`triple--pomeron vertex' which describes the splitting of one BFKL
pomeron into two within the terms responsible for fluctuations. The
ensuing triple--pomeron vertex coincides with the one
previously derived by Braun and Vacca within perturbative QCD.
The evolution equations can be recast in a Langevin form, but with a
multivariable noise term with off--diagonal correlations.
Our equations are shown to be equivalent with the modified version of the 
JIMWLK equation recently proposed by Mueller, Shoshi, and Wong.

\end{abstract}
\end{center}}

\end{frontmatter}
\newpage

\section{Introduction}
\setcounter{equation}{0}

In a recent publication \cite{IT04}, we have shown that the
Balitsky--JIMWLK (Jalilian-Marian--Iancu--McLerran--Weigert--Leonidov--Kovner)
 hierarchy of equations \cite{B,JKLW97,RGE,W}
previously proposed to describe the non--linear evolution of scattering
amplitudes in QCD at high energy fails to include the effects of
fluctuations in the gluon number in the target wavefunction, that were
recently demonstrated to play an important role in the approach towards
saturation and the unitarity limit \cite{MS04,IMM04}. Still in Ref.
\cite{IT04}, we have then constructed a generalization of these
equations which includes the relevant fluctuations in the limit where
the number of colors is large ($N_c\gg 1$) and in the kinematical
region where these effects are expected to be important
\cite{MS04,IMM04}, namely, in the low density regime at high transverse
momenta (well above the target saturation momentum \cite{GLR,MV}). To
that aim, we have relied on Mueller's color dipole picture
\cite{AM94,AM95,Salam95}, which is indeed known to provide a complete
description of the target wavefunction (including the relevant
fluctuations) in the approximation of interest. Besides, in order to
simplify some of the technical manipulations in Ref. \cite{IT04}, we
have performed some further approximations in  the relation between
scattering amplitudes and dipole densities, which are tantamount to
neglecting the non--locality of the two--gluon exchanges in
dipole--dipole scattering.

Although reasonable within the most interesting kinematical region ---
namely, within the intermediate regime at high momenta which is
characterized by the BFKL `anomalous dimension'
\cite{BFKL,GLR,SCALING,MT02,DT02,MP03} ---, these additional approximations have
ineluctably entailed a loss of accuracy, and also some conceptual
shortcomings, for our final results in Ref. \cite{IT04}. Specifically,
the normalization of the fluctuation terms in our final equations there
is afflicted with an unknown `fudge' factor, and the non--locality
structure of those terms is oversimplified. Because of that, these
equations cannot be used, e.g., to study the transition towards the
high--$Q^2$ evolution \cite{DGLAP}, which is expected to take place at
sufficiently large transverse momenta. For instance, one would like to
understand the interplay between the fluctuation--dominated dynamics
predicted by the analysis in Refs. \cite{IMM04,IT04} (which should
completely supersede the BFKL dynamics at very high energies) and the
DGLAP--like dynamics \cite{DGLAP} which should emerge at higher
transverse momenta. But in order to do so, one would need the correct
non--local structure of all the terms in the equations, including the
fluctuation terms. Besides, the too drastic simplifications performed
on the structure of the fluctuation terms leads to numerical
difficulties which prevent a direct resolution of the hierarchy of
equations proposed in Ref. \cite{IT04} through iterations \cite{Anna}.
These difficulties are eliminated by an additional coarse--graining in
impact parameter space, which leads to a mathematically well--defined
Langevin equation  \cite{IT04} --- the QCD analog of the stochastic
Fisher--Kolmogorov--Petrovsky--Piscounov (sFKPP) equation \cite{Saar}.
But this final equation has the further drawback to ignore the
impact--parameter dependence of the scattering amplitudes. Last but not
least, we would like to be able to compare the newly proposed terms in
Ref. \cite{IT04} with related results in perturbative QCD
\cite{BW93,RP97,BV99,ES04,BLV05,BBV05}. As explained in \cite{IT04}, these
terms describe pomeron splittings in the target wavefunction. The
relevant triple--pomeron vertex has been computed in perturbative QCD
in the recent years \cite{BW93,BV99,BLV05,BBV05}, and we would like to
identify this as a building block in our fluctuation terms.

For all these reasons it would be important to relax the additional
approximations performed in Ref. \cite{IT04}, and thus obtain the
complete equations valid to leading order at large $N_c$. This is the
task that we shall undertake in this Letter. Specifically, by using the
complete relations  between scattering amplitudes and dipole densities
valid in the dilute regime, we shall be able to restore the full
non--locality of the fluctuation terms in the evolution equations for
the amplitudes, and thus correct all the deficiencies alluded to
before. In particular, in our new equations we shall recognize, as
expected, the large--$N_c$ version of the triple--pomeron vertex
describing pomeron {\it splitting}, in the form given by Braun and
Vacca \cite{BV99} (see also \cite{RP97}). Since we knew already that
the triple--pomeron vertex responsible for pomeron {\it mergings} is
also included \cite{Braun00,BLV05}, through the non--linear terms in
our equations (which are the same as the corresponding terms in the
Balitsky--JIMWLK hierarchy \cite{B,JKLW97,RGE,W}), we conclude that the
new equations generate correctly, through iterations, the {\it pomeron
loops} expected in perturbative QCD at large $N_c$ \cite{AM95,Salam95}.

In view of numerical studies, it is often convenient to dispose of a
stochastic formulation of a hierarchy of equations. We shall show that
our new equations can be equivalently reformulated as a Langevin
equation, which generalizes the one proposed  in Ref. \cite{IT04} by
including the impact--parameter dependence and the full non--locality
of the coefficient of the noise term. The counterpart of this is the
fact that the new noise term depends upon three variables, and has
non--diagonal correlations in two of them, which rises some doubts
about his usefulness for practical calculations.

In this paper, we shall follow the same general strategy as in Ref.
\cite{IT04}, namely, we shall view the high--energy evolution as the
evolution of the target wavefunction, for which we shall use the dipole
picture \cite{AM94} in the dilute regime at high momenta and the JIMWLK
evolution \cite{JKLW97,RGE,W} in the high density regime near, or at,
saturation. Very recently, Mueller, Shoshi, and Wong  \cite{MSW05} have
proposed a generalization of the JIMWLK equation which includes the
effects of pomeron splitting in the dilute regime, at the same accuracy
level as the dipole picture. We shall explicitly check that our new
equations are indeed equivalent with the evolution described in Ref.
\cite{MSW05}. In particular, the need for a non--diagonal noise term in
order to achieve a stochastic formulation has been originally
recognized in \cite{MSW05}.

The paper is structured as follows: In Sect. 2, we shall construct the
fluctuation term for the equation describing the evolution of the
scattering amplitude for a pair of dipoles (which is the lowest
equation in the hierarchy where this term enters). Then, in Sect. 3, we
shall generalize this construction to the equations of higher level in
the hierarchy (pertinent to the scattering of $\kappa$ dipoles with
$\kappa\ge 3$), and we shall introduce the alternative formulation of
the hierarchy as a Langevin equation. In Sect. 4 we shall demonstrate
the equivalence with the extended JIMWLK formalism of Ref.
\cite{MSW05}. Finally, in Sect. 5 we shall analyze the structure of the
triple--pomeron vertex encoded in our equation, and recognize this as
the vertex originally computed by Braun and Vacca \cite{BV99}.

\section{The scattering amplitude for a pair of dipoles}

In this section we will derive the evolution equation for the amplitude
describing the scattering of a dipole pair off an arbitrary hadronic
target in the large--$N_c$ limit. We shall view the scattering in a
frame in which most of the total rapidity $Y$, with $\bar\alpha_s Y >
1$, is carried by the target (whose wavefunction is therefore highly
evolved), whereas the projectile is simply a set of elementary
dipoles. Also, we shall discuss explicitly only the contributions to
the evolution arising from fluctuations in the gluon number in the
target wavefunction. From the discussion in Refs.
\cite{MS04,IMM04,IT04}, we expect such fluctuations to play an
important role only in the tail of the distribution at large transverse
momenta, where the gluon occupation factor is low (of order one), and
the dipole picture is reliable (for large $N_c$). To study this low
density region, we shall assume that the external dipoles composing the
projectile have small transverse sizes (much smaller than the inverse
of the target saturation momentum at their respective impact
parameters).

Consider first the scattering of a single external dipole, with
transverse coordinates $\bm{x}$ and $\bm{y}$ (for the quark and
antiquark, respectively). In the present approximations, this dipole
undergoes single scattering via two gluon exchange with any of the
dipoles internal to the target. Thus, the corresponding scattering
amplitude reads:
\begin{align}\label{Tconv}
    \mean{T(\bm{x},\bm{y})} = \alpha_s^2
    \int \dif^2 \bm{u}\, \dif^2 \bm{v}\,
    \cal{A}_0(\bm{x},\bm{y}|\bm{u},\bm{v})\,
    n_Y(\bm{u},\bm{v})
\end{align}
where, up to a factor $\alpha_s^2$, $\cal{A}_0$ is the amplitude for
dipole--dipole scattering in the two--gluon exchange approximation and
for large $N_c$ :
\begin{align}\label{T0}
    \cal{A}_0(\bm{x},\bm{y}|\bm{u},\bm{v}) =
    \frac{1}{8}
    \left[\ln \frac{(\bm{x}-\bm{v})^2 (\bm{y}-\bm{u})^2}
    {(\bm{x}-\bm{u})^2 (\bm{y}-\bm{v})^2}
    \right]^2.
\end{align}
The factor of $\alpha_s^2$ has been extracted from the amplitude but
included explicitly in Eq.~(\ref{Tconv}), in order to facilitate the
perturbative counting for the subsequent expressions. Furthermore,
$n_Y(\bm{u},\bm{v})$ measures the average number of dipoles with
coordinates $(\bm{u},\bm{v})$ in the wavefunction of the hadron as
produced after a BFKL evolution over the rapidity $Y$.

At this point, and for later convenience, it is useful to invert
Eq.~(\ref{Tconv}) and express the dipole density in terms of the
scattering amplitude. Applying $\nabla_{\bm{x}}^2 \nabla_{\bm{y}}^2$
on Eq.~(\ref{Tconv}) and using the property $\nabla_{\bm{x}}^2 \ln
\bm{x}^2 = 4 \pi \delta^{(2)}(\bm{x})$, we obtain
\begin{align}\label{invert}
    n_Y(\bm{x},\bm{y}) + n_Y(\bm{y},\bm{x})=
    \frac{4}{g^4}
    \nabla_{\bm{x}}^2 \nabla_{\bm{y}}^2\, \mean{T(\bm{x},\bm{y})},
\end{align}
with the assumption $\bm{x}\neq \bm{y}$.

Now let us consider the scattering of a pair of dipoles
$(\bm{x}_1,\bm{y}_1)$ and $(\bm{x}_2,\bm{y}_2)$ off the target.
Clearly the extension of Eq.~(\ref{Tconv}) reads
\begin{align}\label{T2conv}
    \mean{T^{(2)}(\bm{x}_1,\bm{y}_1;\bm{x}_2,\bm{y}_2)} =
    \alpha_s^4\int\limits_{\bm{u}_i,\bm{v}_i}
    \cal{A}_0(\bm{x}_1,\bm{y}_1|\bm{u}_1,\bm{v}_1)\,
    \cal{A}_0(\bm{x}_2,\bm{y}_2|\bm{u}_2,\bm{v}_2)\,
    n^{(2)}_Y(\bm{u}_1,\bm{v}_1;\bm{u}_2,\bm{v}_2),
\end{align}
where $ n^{(2)}_Y(\bm{u}_1,\bm{v}_1;\bm{u}_2,\bm{v}_2)$ is the average
number density of pairs of dipoles in the target. As in Ref.
\cite{IT04}, we include in  $ n^{(2)}_Y$ only the pairs made of two
{\it different} dipoles, that is, we do not allow the two external
dipoles to scatter off a {\it same} dipole in the target. This should
be a good approximation so long as the external dipoles have different
sizes and/or impact parameters. In particular, if at $Y=0$ the target
consists in a single dipole (this will be our initial condition), then
$n^{(2)}_0=0$, and therefore $\langle T^{(2)}\rangle_0=0$ as well.

The evolution equation for
$\means{T^{(2)}}$ can be obtained from the corresponding equation
for $n^{(2)}_Y$, which has been derived in Ref. \cite{IT04}.
The inhomogeneous part of the latter, i.e. the
piece which describes low--density fluctuations, reads
\begin{align}\label{n2evol}
    \frac{\partial n_Y^{(2)}(\bm{u}_1,\bm{v}_1;\bm{u}_2,\bm{v}_2)}
    {\partial Y}\,
    \bigg|_{\rm fluct} =
    \frac{\abar}{2\pi}\,
    \mathcal{M}(\bm{u}_1,\bm{v}_2,\bm{u}_2)\,
    n_Y(\bm{u}_1,\bm{v}_2)\,
    \delta^{(2)}(\bm{u}_2-\bm{v}_1)\,
    +\, \big\{1\leftrightarrow 2\big\},
\end{align}
with the dipole kernel \cite{AM94}
\begin{equation}\label{dipkernel}
    \mathcal{M}(\bm{x},\bm{y},\bm{z})=
    \frac{(\bm{x}-\bm{y})^2}
    {(\bm{x}-\bm{z})^2 (\bm{y}-\bm{z})^2}\,.
\end{equation}
(The terms not shown here correspond to the usual BFKL evolution of the
dipole pair density; see Eq.~(5.15) in Ref. \cite{IT04}.)
Using Eqs.~(\ref{T2conv}) and (\ref{n2evol}) to write the evolution
of $\means{T^{(2)}}$ in terms of the dipole density $n_Y$ and then
employing Eq.~(\ref{invert}) in order to express $n_Y$ in terms of
the scattering amplitude $\means{T}$, it is straightforward to show
that
\begin{align}\label{T2evol}
    \frac{\partial \mean{T^{(2)}(\bm{x}_1,\bm{y}_1;\bm{x}_2,\bm{y}_2)}}
    {\partial Y}\,
    \bigg|_{\rm fluct} =
    \left(\frac{\alpha_s}{2\pi}\right)^2
    \frac{\abar}{2 \pi}\!
    \int\limits_{\bm{u},\bm{v},\bm{z}}&
    \mathcal{M}(\bm{u},\bm{v},\bm{z})\,
    \cal{A}_0(\bm{x}_1,\bm{y}_1|\bm{u},\bm{z})\,
    \cal{A}_0(\bm{x}_2,\bm{y}_2|\bm{z},\bm{v})\,
    \nonumber \\
    &\times\nabla_{\bm{u}}^2 \nabla_{\bm{v}}^2\, \mean{T(\bm{u},\bm{v})}.
\end{align}

The complete evolution equation for  $\means{T^{(2)}}$ can be now
obtained by restoring the more standard terms associated with the BFKL
evolution and with unitarity corrections at high energy.
The former are terms linear in $\means{T^{(2)}}$, while the latter
are non--linear terms
proportional to the scattering amplitude $\means{T^{(3)}}$ for a system
of three external dipoles, and are the same as in the corresponding 
Balitsky--JIMWLK equation \cite{B,RGE,W} (here, for large $N_c$). 
Thus our final evolution equation for $\means{T^{(2)}}$ reads
\begin{align}\label{T2evolfull}
    \frac{\partial \mean{T^{(2)}(\bm{x}_1,\bm{y}_1;\bm{x}_2,\bm{y}_2)}}
    {\partial Y}\,=
    \frac{\abar}{2\pi}& \int\limits_{\bm{z}}
    \big\{\big[\cal{M}_{\bm{x}_1\bm{y}_1\bm{z}}
    \otimes \mean{T^{(2)}(\bm{x}_1,\bm{y}_1;\bm{x}_2,\bm{y}_2)}
    \nonumber \\
    &-\cal{M}(\bm{x}_1,\bm{y}_1,\bm{z})\,
    \mean{T^{(3)}(\bm{x}_1,\bm{z};\bm{z},\bm{y}_1;\bm{x}_2,\bm{y}_2)}\big]
    + [1\leftrightarrow 2]\big\}
    \nonumber \\
    &+\frac{\partial \mean{T^{(2)}(\bm{x}_1,\bm{y}_1;\bm{x}_2,\bm{y}_2)}}
    {\partial Y}\,
    \bigg|_{\rm fluct},
\end{align}
with the last term given by Eq.~(\ref{T2evol}) and where we have
introduced the shorthand notation
\begin{align}\label{Moper}
    \cal{M}_{\bm{x}\bm{y}\bm{z}}
    \otimes f(\bm{x},\bm{y}) \equiv \cal{M}(\bm{x},\bm{y},\bm{z})
    [-f(\bm{x},\bm{y})+f(\bm{x},\bm{z})+f(\bm{z},\bm{y}) ].
\end{align}
Under the same approximations, the scattering amplitude for a single
dipole obeys the following equation, which is formally the same as the
first equation in Balitsky hierarchy:
 \begin{align}\label{T1evolfull}
    \frac{\partial \mean{T(\bm{x},\bm{y})}}
    {\partial Y}\,=
    \frac{\abar}{2\pi} \int\limits_{\bm{z}}
    \big\{\cal{M}_{\bm{x}\bm{y}\bm{z}}
    \otimes \mean{T(\bm{x},\bm{y}}
       &-\cal{M}(\bm{x},\bm{y},\bm{z})\,
    \mean{T^{(2)}(\bm{x},\bm{z};\bm{z},\bm{y})}\big\}.
\end{align}
Note that the evolutions of $\mean{T}$ and $\means{T^{(2)}}$ are
coupled by the fluctuation term in Eq.~(\ref{T2evolfull}), in agreement
with \cite{IT04}. Moreover, $\means{T^{(2)}}$ is also coupled to
$\means{T^{(3)}}$, by the unitarity corrections, so the equations above
do not close by themselves, but rather are a part of an extended
(actually, infinite) hierarchy. The higher equations in this hierarchy
will be written down in the next section.

Before closing this section, we would like to comment on the
simplified version of Eq.~(\ref{T2evol}) which was proposed in
\cite{IT04}. There, an approximation was done at the level of
Eq.~(\ref{T0}), in which the non--locality of the 2--gluon exchange
interaction was neglected. That is, the elementary dipole--dipole scattering
amplitude was effectively assumed to be
\begin{align}\label{T0approx}
    \alpha_s^2
    \cal{A}_0(\bm{x},\bm{y}|\bm{u},\bm{v}) \simeq
    c\,\alpha_s^2
    (\bm{x}-\bm{y})^4\,
    \delta^{(2)}(\bm{x}-\bm{u})\,
    \delta^{(2)}(\bm{y}-\bm{v}),
\end{align}
where $c$ is an unknown ``fudge'' factor. Then, by following the
same steps as above, it was shown that the right hand side of
Eq.~(\ref{T2evol}) is
\begin{align}\label{T2approx}
    c\,\alpha_s^2\,
    \frac{\abar}{2\pi}\,
    \frac{(\bm{x}_1-\bm{y}_1)^2 (\bm{x}_2-\bm{y}_2)^2}
    {(\bm{x}_1+\bm{y}_2)^2}\,
    \mean{T({\bm x}_1,{\bm y}_2)}\,
    \delta^{(2)}({\bm x}_2-{\bm y}_1)
    \,+\,\big\{1 \leftrightarrow 2 \big\}.
\end{align}
After a coarse--graining in impact parameter space and a saddle point
approximation, a Langevin equation, equivalent to the stochastic FKPP
equation, was finally obtained. However, these equations suffer from
the shortcomings mentioned in the Introduction.

\section{The Langevin Equation}
\setcounter{equation}{0}

In order to have a complete hierarchy of equations, we need to generalize
Eq.~(\ref{T2evol}) to the case where an
arbitrary number of dipoles, say $\kappa$, scatters of the target.
Given the $\kappa$-th density of dipoles
$n_Y(\bm{u}_1,\bm{v}_1;...;\bm{u}_{\kappa},\bm{v}_{\kappa}) \equiv
n_Y(\{\bm{u},\bm{v}\})$ in the wavefunction of the hadron, the amplitude
for $\kappa$-dipoles scattering reads
\begin{align}\label{Tkconv}
    \mean{T^{(\kappa)}(\{\bm{x},\bm{y}\})} =
    \alpha_s^{2\kappa}\!
    \int\limits_{\bm{u}_i,\bm{v}_i}
    \cal{A}_0(\bm{x}_1,\bm{y}_1|\bm{u}_1,\bm{v}_1)
    \,...\,\cal{A}_0(\bm{x}_{\kappa},\bm{y}_{\kappa}|
    \bm{u}_{\kappa},\bm{v}_{\kappa})\,
    n^{(\kappa)}_Y(\{\bm{u},\bm{v}\}).
\end{align}
The corresponding evolution equation is obtained by the straightforward
generalization of the procedure that we followed in the previous section.
The rate of change, due to fluctuations, of the $\kappa$-th density
of dipoles of the hadronic wavefunction depends only in the
$(\kappa-1)$-th density, and therefore the same will be true for the
scattering amplitude $\means{T^{(\kappa)}(\{\bm{x},\bm{y}\})}$,
since Eq.~(\ref{Tkconv}) relates quantities of the ``same order'' in
$\kappa$. The ensuing equation reads
\begin{align}\label{Tkevol}
    \frac{\partial \mean{T^{(\kappa)}(\{\bm{x},\bm{y}\})}}
    {\partial Y}\,
    \bigg|_{\rm fluct}=
    \left(\frac{\alpha_s}{2\pi}\right)^2
    \frac{\abar}{2 \pi}\!
    &\int\limits_{\bm{u},\bm{v},\bm{z}}
    \mathcal{M}(\bm{u},\bm{v},\bm{z})\,
    \cal{A}_0(\bm{x}_1,\bm{y}_1|\bm{u},\bm{z})\,
    \cal{A}_0(\bm{x}_2,\bm{y}_2|\bm{z},\bm{v})\,
    \nonumber \\
    &\times \nabla_{\bm{u}}^2 \nabla_{\bm{v}}^2\, \mean{T^{(\kappa-1)}
    (\bm{u},\bm{v};\bm{x}_3,\bm{y}_3;...;\bm{x}_{\kappa},\bm{y}_{\kappa})}
    \, +\,{\rm perm.},
\end{align}
where ``perm.'' stands for all possible permutations in the arguments of
$\means{T^{(\kappa-1)}}$, so that the total number of terms in the
right hand side is $\kappa (\kappa-1)/2$.

It is now straightforward to add to the r.h.s. of Eq.~(\ref{Tkevol})
the (large--$N_c$ version of the) standard Balitsky--JIMWLK terms, and thus obtain the complete 
hierarchy for $\means{T^{(\kappa)}}$ within the present approximation.
Since in practice (e.g., for numerical studies) it seems difficult to deal with a hierarchy
as a whole, it is convenient to notice that one can equivalently reformulate this hierarchy
as a single stochastic equation. Let us consider the following Langevin equation
\begin{align}\label{Tlang}
    \frac{\partial T_Y(\bm{x},\bm{y})}
    {\partial Y}\,
    \Big|_{\rm fluct}=
    \frac{\alpha_s}{2\pi}\,
    \sqrt{\frac{\abar}{2 \pi}}\!
    \int\limits_{\bm{u},\bm{v},\bm{z}}
    \cal{A}_0(\bm{x},\bm{y}|\bm{u},\bm{z})
    \frac{|\bm{u}-\bm{v}|}{(\bm{u}-\bm{z})^2}
    \sqrt{\nabla_{\bm{u}}^2 \nabla_{\bm{v}}^2\, T_Y(\bm{u},\bm{v})}\,
    \nu(\bm{u},\bm{v},\bm{z},Y)\,,
\end{align}
where the noise satisfies
\begin{align}\label{noise}
    \langle \nu(\bm{u}_1,\bm{v}_1,\bm{z}_1,Y)
    \nu(\bm{u}_2,\bm{v}_2,\bm{z}_2,Y') \rangle=
    \delta^{(2)}(\bm{u}_1-\bm{v}_2)
    \delta^{(2)}(\bm{v}_1-\bm{u}_2)
    \delta^{(2)}(\bm{z}_1-\bm{z}_2)
    \delta(Y-Y')
\end{align}
with all the other correlators being zero. Note that the correlator (\ref{noise})
is {\it non--diagonal} in the first two arguments $(\bm{u},\bm{v})$
of the noise. A similar property has been recently noticed in Ref. \cite{MSW05},
in a physical context which is very close to ours (see the discussion in Sect. 4 below),
but we are not aware about previous studies of such a non--diagonal noise in other
(physical or mathematical) contexts. The stochastic problem above
must be understood with the Ito prescription for rapidity discretization
(see, e.g., \cite{IT04,PATH} for details). Then one can check that by 
solving this equation one generates the same correlations as from
the original hierarchy in Eq.~(\ref{Tkevol}). 
Indeed, consider the change in the $\kappa$--dipole function
$T_1 ... T_{\kappa}$ under an increase $\Delta Y$ in rapidity
(in $T_i$, the index $i$ serves as a shorthand
for the coordinates of the $i$-th dipole). Eqs.~(\ref{Tlang}) and
(\ref{noise}) imply that $\Delta T \sim \sqrt{\Delta Y}$, so in
evaluating $ \Delta (T_1...T_{\kappa})$, it is enough to keep terms
of second order in the $\Delta T$'s :
\begin{align}\label{delta}
    \Delta (T_1...T_{\kappa})=
    (T_1 + \Delta T_1)...
    (T_{\kappa} + \Delta T_{\kappa})-
    T_1 ... T_{\kappa} \rightarrow
    \Delta T_1\, \Delta T_2\,
    T_3\, ... \,T_{\kappa}
    + {\rm perm.}
\end{align}
Indeed, the linear terms will vanish after averaging over the noise,
whereas the cubic terms, or higher, will be of higher order in $\Delta Y$,
and thus will vanish in the continuum limit $\Delta Y\to 0$. The
transverse coordinates dependence of the noise correlation in
Eq.~(\ref{noise}) is such that, when we take the expectation value in
Eq.~(\ref{delta}), the emerging hierarchy is exactly the same as the
one in Eq.~(\ref{Tkevol}), provided we make the identification
$\langle T_1...T_{\kappa} \rangle \leftrightarrow
\means{T^{(\kappa)}(\{\bm{x},\bm{y}\})}$.

Eq.~(\ref{Tlang}) refers solely to the fluctuations terms in the hierarchy,
but the standard Balitsky--JIMWLK terms can be easily added \cite{IT04}. 
We finally obtain 
\begin{align}\label{Tlangfull}
    \frac{\partial T_Y(\bm{x},\bm{y})}
    {\partial Y}\, =\,&
    \frac{\abar}{2\pi} \int\limits_{\bm{z}}
    \big[\cal{M}_{\bm{x}\bm{y}\bm{z}}
    \otimes T_Y(\bm{x},\bm{y})
    -\cal{M}(\bm{x},\bm{y},\bm{z})\,
    T_Y(\bm{x},\bm{z})\,T_Y(\bm{z},\bm{y})\big]
    \nonumber\\
    &+\frac{\partial T_Y(\bm{x},\bm{y})}{\partial Y}\,
    \Big|_{\rm fluct},
\end{align}
with the last term being given by Eq.~(\ref{Tlang}). This equation
together with the noise term (\ref{noise}) is the Langevin
problem equivalent to the non--linear evolution equations with
pomeron loops. But, as mentioned above, the non--diagonal nature of
the noise correlator (\ref{noise}) is a rather unusual mathematical
property, and presently
we do not know whether such a noise can be useful for numerical studies, or not.

\section{The Color Glass Condensate Approach}
\setcounter{equation}{0}

An alternative approach to high--energy QCD is the
Color Glass Condensate (CGC) formalism in which one studies
the random color fields created by `color sources' (gluons or dipoles)
within the target, and the evolution
of the respective correlation functions with increasing $Y$ (see
\cite{CGCreviews} for review papers and more references).
In a specific gauge, the color field in the target has only
one component, the light-cone component $A^+_a$, which is traditionally
denoted as $A^+_a(x)\equiv\alpha^a(x^-,\bm{x})$. (This is a function
of the longitudinal coordinate $x^-$ together with the transverse
coordinate $\bm{x}$.) Let $W_Y[\alpha]$ represent the functional probability 
distribution for finding a particular configuration of this field at rapidity $Y$. 
Then the expectation value of an operator
$\cal{O}[\alpha]$ is given by
\begin{align}\label{Omean}
    \mean{\cal{O}[\alpha]}=\int D\alpha\, W_Y[\alpha]
    \cal{O}[\alpha].
\end{align}
For example the operator for the scattering between a single dipole
$(\bm{x},\bm{y})$ and the CGC is computed in the eikonal approximation as
\begin{align}\label{Twil}
    T(\bm{x},\bm{y})=1-\frac{1}{N_c}\,
    {\rm tr} \left( V_{\bm{x}}^\dagger V_{\bm{y}}\right),
\end{align}
where $V_{\bm{x}}^\dagger$ is a Wilson line describing the 
scattering between the quark and the color field in the CGC
(below, $\cal{P}$ stands for a path ordered product),
\begin{align}\label{wilson}
    V_{\bm{x}}^\dagger[\alpha]= \cal{P}\exp \left[
    i g \int \dif x^- \alpha^a(x^-,\bm{x}) t^a \right],
\end{align}
while $V_{\bm{y}}$ refers similarly to the antiquark. Furthermore, the
corresponding operator for a dipole pair is
$T^{(2)}(\bm{x}_1,\bm{y}_1;\bm{x}_2,\bm{y}_2) =
T(\bm{x}_1,\bm{y}_1)T(\bm{x}_1,\bm{y}_1)$, with $T$ given by
Eq.~(\ref{Twil}). 

Under an increase $\dif Y$ in rapidity, the correlation functions of the
color field change because of the induced emission of a 
small--$x$ gluon. These changes can be accommodated in a functional 
evolution equation for the probability distribution, 
which can be written in Hamiltonian form (at least, formally) :
\begin{align}\label{Wevol}
    \frac{\partial}{\partial Y} W_Y[\alpha]=
    H\, W_Y[\alpha].
\end{align}
Then the evolution equation for $\mean{\cal{O}[\alpha]}$ can be deduced
from Eqs.~(\ref{Omean}) and (\ref{Wevol}).

The original version of the Hamiltonian which enters Eq.~(\ref{Wevol}), known
as the JIMWLK Hamiltonian ($H_{\rm JIMWLK}$) \cite{JKLW97,RGE,W}, leads to Balitsky 
equations \cite{B} when applied to scattering operators like (\ref{Twil}). 
$H_{\rm JIMWLK}$ is a second--order functional differential operator,
with a kernel which is non--linear in the field $\alpha$ to all orders.
The quadratic part of this kernel is responsible for the linear, BFKL, evolution, 
while the terms of higher order in $\alpha$ generate non--linear terms, 
so like the terms expressing unitarity corrections in Balitsky 
equations. As we have noticed in \cite{IT04}, the evolution generated
by $H_{\rm JIMWLK}$ does not include gluon  {\it splitting}, and therefore cannot
describe the correlations associated with fluctuations in the gluon number, which
are however important in the dilute regime. Very recently, Mueller, Shoshi and Wong 
\cite{MSW05} have proposed an additional term $H_{\rm fluct}$, which is inspired
by the probabilistic interpretation of the evolution in the dipole model
\cite{AM94,IM031}, and which takes into account the low density fluctuations
at the same level of accuracy as the dipole picture. Thus, the total Hamiltonian
used in Ref. \cite{MSW05} can be written as:
\begin{align}\label{Hamsep}
    H=H_{\rm JIMWLK} + H_{\rm fluct},
\end{align}
where $H_{\rm JIMWLK}$ can be found in Refs. \cite{RGE,W} (see also \cite{ODDERON}
for a simple expression), and
\begin{align}\label{Ham}
    H_{\rm fluct} =
    -\frac{\pi \alpha_s}{4 N_c^3}\,
    \frac{\abar}{2 \pi}
    \int & \cal{M}(\bm{u},\bm{v},\bm{z})
    \cal{G}(\bm{u}_1|\bm{u},\bm{z})
     \cal{G}(\bm{v}_1|\bm{u},\bm{z})
    \cal{G}(\bm{u}_2|\bm{z},\bm{v})
    \cal{G}(\bm{v}_2|\bm{z},\bm{v})
    \nonumber\\
    & \times
    \frac{\delta}{\delta \alpha^a(\bm{u}_1)}
    \frac{\delta}{\delta \alpha^a(\bm{v}_1)}
    \frac{\delta}{\delta \alpha^b(\bm{u}_2)}
    \frac{\delta}{\delta \alpha^b(\bm{v}_2)}
    \nabla_{\bm{u}}^2 \nabla_{\bm{v}}^2
    \alpha^c(\bm{u}) \alpha^c(\bm{v}).
\end{align}
Here, the integration goes over all the transverse coordinates
$\bm{u},\bm{v},\bm{z},\bm{u}_1,\bm{v}_1,\bm{u}_2,\bm{v}_2$, and
$\alpha^{a}(\bm{x})\equiv \int dx^- \alpha^a(x^-,\bm{x})$.
The function $\cal{G}(\bm{u}_1|\bm{u},\bm{z})$ is, up to
a factor $g \,t^a$, the classical field created by the elementary
dipole $(\bm{u},\bm{z})$, and reads
\begin{align}\label{calG}
    \cal{G}(\bm{u}_1|\bm{u},\bm{z}) =
    \frac{1}{4\pi}
    \ln \frac{(\bm{u}_1-\bm{z})^2}{(\bm{u}_1-\bm{u})^2}.
\end{align}

In what follows we would like to show that the Hamiltonian (\ref{Ham}) leads 
precisely to
our previous hierarchy (\ref{Tkevol}) for the dipole scattering amplitudes. 
We shall demonstrate that only for the case of two external dipoles
($\kappa=2$), since the corresponding proof for arbitrary $\kappa$ follows
trivially. Also we shall focus on the contribution
induced by the fluctuation piece  $H_{\rm fluct}$ of the Hamiltonian,
since the corresponding contribution of $H_{\rm JIMWLK}$ is well--known
\cite{RGE,W,PATH,ODDERON} to be the respective Balitsky equation.
Thus, we start with (cf. Eqs.~(\ref{Omean}) and (\ref{Wevol})):
\begin{align}\label{EvolT2}
    \frac{\partial \mean{T^{(2)}(\bm{x}_1,\bm{y}_1;\bm{x}_2,\bm{y}_2)}}
    {\partial Y}\,
    \bigg|_{\rm fluct} =
    \int D \alpha\,
    T^{(2)}(\bm{x}_1,\bm{y}_1;\bm{x}_2,\bm{y}_2)\,
    H_{\rm fluct}\, W_Y[\alpha].
\end{align}
In the weak field limit the scattering amplitude is determined from
the small coupling expansion of the corresponding Wilson lines
operator and it reads
\begin{align}\label{T2alpha}
    T^{(2)}(\bm{x}_1,\bm{y}_1;\bm{x}_2,\bm{y}_2)=
    \frac{g^4}{16 N_c^2}\,
    [\alpha^{d}(\bm{x}_1) - \alpha^{d}(\bm{y}_1)]^2
    [\alpha^{e}(\bm{x}_2) - \alpha^{e}(\bm{y}_2)]^2.
\end{align}
By partial integrations within the r.h.s. of Eq.~(\ref{EvolT2}), one can bring 
the functional differential operator $(\delta/\delta\alpha)^4$ (cf. Eq.~(\ref{Ham}))
to act on $T^{(2)}$. This operation annihilates the
$\alpha$ fields manifest in Eq.~(\ref{T2alpha})
and gives rise to a sum of products of two--dimensional delta functions. 
For consistency with the derivation of Eq.~(\ref{Ham}) \cite{MSW05}, we should
restrict ourselves to the terms of leading order in $N_c$, which
arise when a pair of derivatives which are traced over color
acts on a pair of fields which carry both the same color indices. Then
we can perform the integrations over
$\bm{u}_1,\bm{v}_1,\bm{u}_2,\bm{v}_2$, to finally arrive at 
\begin{align}\label{T2evolhamtemp}
    \frac{\partial \mean{T^{(2)}(\bm{x}_1,\bm{y}_1;\bm{x}_2,\bm{y}_2)}}
    {\partial Y}\,
    \bigg|_{\rm fluct} =
    \frac{\alpha_s^3}{4 \pi N_c}
    \frac{\abar}{2\pi}
    \int\limits_{\bm{u},\bm{v},\bm{z}}&
    \mathcal{M}(\bm{u},\bm{v},\bm{z})\,
    \cal{A}_0(\bm{x}_1,\bm{y}_1|\bm{u},\bm{z})\,
    \cal{A}_0(\bm{x}_2,\bm{y}_2|\bm{z},\bm{v})\,
    \nonumber \\
    &\times\nabla_{\bm{u}}^2 \nabla_{\bm{v}}^2\,
    \mean{-2\alpha^c(\bm{u}) \alpha^c(\bm{v})}.
\end{align}
The last factor in the above equation may be rewritten as
\begin{align}
    \nabla_{\bm{u}}^2 \nabla_{\bm{v}}^2\,
    \mean{-2\alpha^c(\bm{u}) \alpha^c(\bm{v})} &=
   \nabla_{\bm{u}}^2 \nabla_{\bm{v}}^2\,
    \mean{[\alpha^c(\bm{u})-\alpha^c(\bm{v})]^2}=
    \frac{N_c}{\pi \alpha_s}\,
    \nabla_{\bm{u}}^2 \nabla_{\bm{v}}^2\,
    \mean{T(\bm{u},\bm{v})}
    ,
\end{align}
where we have also used the weak--field expansion of the scattering amplitude
(\ref{Twil}) for a single dipole. With this identification, Eq.~(\ref{T2evolhamtemp}) 
leads indeed to Eq.~(\ref{T2evol}), which completes our proof.

\section{The triple pomeron vertex}
\setcounter{equation}{0}

The above derivation of our new equations within the CGC approach 
sheds more light on their physical interpretation and confirms
the original discussion in Ref. \cite{IT04}:
when the evolution is included in the target wavefunction,
the {\it fluctuation terms} in the equations for the amplitudes
(e.g., the term linear in 
$\means{T^{(\kappa-1)}}$ in the r.h.s. of  Eq.~(\ref{Tkevol}) for
$\means{T^{(\kappa)}}$) correspond to {\it gluon splittings}, by which we 
mean more precisely the transition vertex from two to four gluons
described by the new piece $H_{\rm fluct}$,  Eq.~(\ref{Ham}), of the CGC Hamiltonian.
Similarly, the {\it  non--linear terms} responsible for unitarization (e.g.,
the term $\means{T^{(3)}}$ in the r.h.s. of  Eq.~(\ref{T2evolfull}) for $\means{T^{(2)}}$),
correspond to {\it gluon mergings}, as described by the terms of higher order
in $\alpha$ in the original JIMWLK Hamiltonian. By `gluon mergings' we mean all
the vertices leading to a reduction in the number of gluon fields. The simplest among 
them is the transition from four to two gluons, and is described by that term in 
$H_{\rm JIMWLK}$ which is {\it dual} to Eq.~(\ref{Ham}), in the sense of 
containing four powers of
$\alpha$ and two powers of the derivative $(\delta/\delta\alpha)$.
In general, one would expect this duality to extend to higher vertices;
that is, $H_{\rm fluct}$ is expected to include terms with higher--order 
derivatives, which should be `dual' to the terms with higher powers of $\alpha$ 
in $H_{\rm JIMWLK}$. Since such terms do not appear to be generated by the dipole
picture which has been used to study fluctuations 
so far \cite{MS04,IMM04,IT04,MSW05}, 
they are presumably of higher order in $1/N_c$.

Furthermore, within the framework of the large--$N_c$ approximation, the
gluon number changing transitions alluded to above should be tantamount
to {\it pomeron splittings}, or {\it mergings}, i.e., processes through
which one BFKL pomeron is connected to two. 
For the non--linear term in the Balitsky--Kovchegov
equation \cite{B,K} (and thus, by extension, for the whole Balitsky hierarchy 
at large $N_c$), this has been checked already in Refs. 
\cite{Braun00,BLV05}. Namely, it has been shown there
that the general 2--to--4 gluon vertex \cite{BW93} reduces essentially to the
dipole kernel,  Eq.~(\ref{dipkernel}), when restricted to the large--$N_c$ limit 
and to the physical conditions appropriate for the merging of two
pomerons\footnote{Namely, the four gluons on one side of the vertex
are pairwise connected in color singlet states, which moreover are multiplied
by functions which vanish at equal coordinates, so like 
$T_Y(\bm{x},\bm{z})\,T_Y(\bm{z},\bm{y})$ in the r.h.s. of 
Eq.~(\ref{Tlangfull}).}. This is indeed consistent with the structure
of the non--linear term in Eq.~(\ref{Tlangfull}).

In what follows, we shall present a similar check for the vertex
that enters the fluctuation term in Eq.~(\ref{T2evol}) (and thus the noise
term in Eq.~(\ref{Tlangfull})).
To that aim, we shall compare our result for this vertex with that in 
Ref. \cite{BV99}, where a linear evolution equation has been constructed which 
includes the triple--pomeron vertex for one pomeron splitting into two. (The 
equation written in Ref. \cite{BV99} refers to the diffractive amplitude for four 
gluons, which is quite similar to the scattering amplitude for two dipoles in our
approach.) 

From our equations, the amputated version of the fluctuation
vertex can be immediately extracted,
as the function which is convoluted with $\means{T(\bm{u},\bm{v})}$ in the r.h.s. 
of Eq.~(\ref{T2evol}). But since in Ref. \cite{BV99} one rather finds the
{\it non--amputated} version of the triple--pomeron vertex (this
is referred there as the `triple pomeron interaction', or TPI), we find it convenient
to compute the similar quantity in our approach as well, 
by solving the linearized version of Eq.~(\ref{T2evolfull}). 
Assuming the initial condition $\langle T^{(2)} \rangle_0=0$, the solution to this
inhomogeneous equation can be written as
\begin{align}\label{T2Green}
    \mean{T^{(2)}(\bm{x}_1,\bm{y}_1;\bm{x}_2,\bm{y}_2)}=
    \int\limits_0^Y \dif y
    \int\limits_{\bm{u}_i,\bm{v}_i}&
    G_{Y-y}(\bm{x}_1,\bm{y}_1|\bm{u}_1,\bm{v}_1)\,
    G_{Y-y}(\bm{x}_2,\bm{y}_2|\bm{u}_2,\bm{v}_2)
    \nonumber \\
    &\times
    \frac{\partial
    \left\langle
    T^{(2)}(\bm{u}_1,\bm{v}_1;\bm{u}_2,\bm{v}_2)
    \right \rangle_y}
    {\partial y}\,
    \bigg|_{\rm fluct},
\end{align}
where the Green's function $G_Y$ satisfies the BFKL equation (which is
homogeneous)
\begin{align}\label{Gevol}
    \left[\frac{\partial}{\partial Y}\,-
    \frac{\abar}{2\pi} \int\limits_{\bm{z}}
    \cal{M}_{\bm{x}\bm{y}\bm{z}}\, \otimes \right]
    G_Y(\bm{x},\bm{y}|\bm{u},\bm{v})=0,
\end{align}
along with the initial condition
\begin{align}\label{Gbound}
    G_Y(\bm{x},\bm{y}|\bm{u},\bm{v})=
    \delta^{(2)}(\bm{x}-\bm{u})
    \delta^{(2)}(\bm{y}-\bm{v}).
\end{align}
Instead of directly solving Eqs.~(\ref{Gevol}) and (\ref{Gbound}),
which is a simple task in any case, it is more convenient for our 
purposes to observe that one can combine each of the Green's function
in Eq.~(\ref{T2Green}) with one of the elementary amplitudes
$\cal{A}_0$ contained in the source term $\partial \langle
T^{(2)}\rangle_y/\partial y$, cf. Eq.~(\ref{T2evol}).
By doing this, we can reconstruct the dipole--dipole
scattering amplitude in the BFKL approximation (the particular
case of the amplitude $\mean{T(\bm{x},\bm{y})}$ in which the target
is itself an elementary dipole).  Indeed, since the dipole--dipole
scattering amplitude in the linear regime satisfies the BFKL
equation, one can write
\begin{align}\label{TGreen}
    \cal{A}_{Y-y}(\bm{x}_1,\bm{y}_1|\bm{u},\bm{z})=
    \int\limits_{\bm{u}_1,\bm{v}_1}
    G_{Y-y}(\bm{x}_1,\bm{y}_1|\bm{u}_1,\bm{v}_1)
    \cal{A}_{0}(\bm{u}_1,\bm{y}_1|\bm{u},\bm{z}),
\end{align}
along with an analogous equation for the $\bm{u}_2,\bm{v}_2$
integration. Therefore, we can finally express the
scattering amplitude for the dipole pair in the form
\begin{align}\label{BVform}
    \mean{T^{(2)}(\bm{x}_1,\bm{y}_1;\bm{x}_2,\bm{y}_2)}=
    \left(\frac{\alpha_s}{2\pi}\right)^2
    \frac{\abar}{2 \pi}\,
    \int\limits_0^Y \dif y
    &\int\limits_{\bm{u},\bm{v},\bm{z}}
    \mathcal{M}(\bm{u},\bm{v},\bm{z})\,
    \cal{A}_{Y-y}(\bm{x}_1,\bm{y}_1|\bm{u},\bm{z})\,
    \nonumber \\
    &\times\cal{A}_{Y-y}(\bm{x}_2,\bm{y}_2|\bm{z},\bm{v})\,
    \nabla_{\bm{u}}^2 \nabla_{\bm{v}}^2\,
    \langle T(\bm{u},\bm{v}) \rangle_y,
\end{align}
which is indeed recognized as the `triple pomeron interaction' in \cite{BV99}
(see Eq. (69) there).
This concludes our present analysis.

\vspace*{0.8cm}

Only a few days ago, when this work was already completed, a preprint appeared
by E.~Levin and M.~Lublinsky \cite{LL05} in which a hierarchy of equations including
pomeron loops is proposed. Although the general structure of this 
hierarchy bears some similarity to ours, its derivation and detailed form 
remain different.

\section*{Acknowledgments}

We would like to thank our colleagues Yoshitaka Hatta,
Larry McLerran, Al Mueller, Arif Shoshi, Anna Stasto and Stephen Wong for
many useful, and critical, comments on our previous preprint
which have inspired some of the developments presented here.

\end{document}